\ifx\empty\selectedoptions
  \def\selectedoptions{final}
\fi

\documentclass[
    a4paper 
    ] {aipproc}

\newcommand{\be}{\begin{equation}} 
\newcommand{\en}{\end{equation}}
\newcommand{\bea}{\begin{eqnarray}}
\newcommand{\ena}{\end{eqnarray}}

\newcommand{\hbo}{\hbox to 1 true cm {\hfill } } 
\newcommand{\tr}{\hbox{tr}}

\layoutstyle{6x9}

\begin{document}

\title 
      [BEC]
      {Confinement versus Bose-Einstein condensation }


\author{Kurt Langfeld}{
  address={Institut f\"ur Theoretische Physik, Universit\"at T\"ubingen, 
  Germany.}
}

\begin{abstract}
The deconfinement phase transition at high baryon densities and 
low temperatures evades a direct investigation by means of lattice 
gauge calculations. In order to make this regime of QCD accessible 
by computer simulations, two proposal are made: (i) A Lattice 
Effective Theory (LET) is designed which incorporates gluon and 
diquark fields. The deconfinement transition takes place when 
the diquark fields undergo Bose-Einstein condensation. 
(ii) Rather than using eigenstates of the particle number operator, 
I propose to perform simulations for a fixed expectation value 
of the baryonic Noether current. This approach changes the view 
onto the finite density regime, but evades the sign and overlap 
problems. The latter proposal is exemplified for the LET: Although the 
transition from the confinement to the condensate phase is 
first order in the coupling constant space at zero baryon densities, 
the transition at finite densities appears to be a crossover. 

\end{abstract}

\date{\today}

\maketitle

\section{Introduction}
Lattice gauge simulations leave no doubt that Quantum Chromodynamics 
(QCD) exhibits a transition from the baryonic regime to the Quark 
Gluon Plasma (QGP) at high temperatures and small baryon densities. 
It is this high temperature regime which is currently under 
investigation at RHIC, Brookhaven~\cite{Heinz:2002gs} and which will be a 
major target of LHC, Cern. It is believed that a transition to the 
QGP also appears at high baryon densities and small temperatures. 
Very little is known about the latter transition from first 
principle simulations: lattice simulations at finite values of the 
baryonic chemical potential $\mu $ encounter a severe sign/overlap 
problem which limits their scope to the range of small 
$\mu $~\cite{deForcrand:2002ci,Fodor:2001au}. 

\vskip 0.3cm 
At asymptotic baryon densities it is assumed that the quarks form a Fermi 
surface. In this case, perturbative gluon interactions support the 
existence of a diquark BCS state known as color 
superconductor~\cite{Bailin:1983bm,Alford:1998mk}. At the present stage 
of research, no first principle results are available for the 
region of the QCD phase diagram where the transition from the 
baryonic phase to the QGP occurs at small temperatures and intermediate 
densities. Here, I will argue that the transition is driven 
by the Bose-Einstein condensation of diquarks. At high densities, 
the Bose-Einstein condensate gradually develops to a diquark 
BCS state. 

\vskip 0.3cm 
In the present paper, two proposal are put forward to provide access 
to the finite density transition of QCD: (i) It is argued that the 
transition is within the reach of a Lattice Effective Theory (LET) which 
incorporates gluons and diquarks as dynamical degrees of freedom. 
At the stage of the present model, the baryonic current is entirely 
supported by the diquarks. 
(ii) As in the case of QCD, the LET also suffers from a sign problem at 
finite baryon chemical potential. In order to get first insights, 
I propose to change the point of view: Rather than considering only 
eigenstates of the particle density operator, simulations are 
performed for a given expectation value of the baryonic Noether 
density. In the case of the LET, we will find that the 
finite density transition is a crossover rather than of 
first order.

\section{Lattice Effective Theory } 

\subsection{Model building } 

The central assumption for describing the finite density transition is 
that only gluon and diquark degrees of freedom are relevant for the 
intermediate density 
region of the phase diagram. In the hadronic phase, diquarks are 
confined to a length scale of $\approx 1 \,$fm. Even if the transition 
is first order, the correlation length might become much larger 
than $\approx 1 \,$fm before the system is disordered by 
bubble nucleation. Therefore, the working hypothesis of the 
present approach is that the degrees of freedom relevant at the 
transition are gluons and point-like scalar (diquark) fields, i.e., 
\be 
\phi ^a (x) \; = \; \epsilon ^{abc} \; \epsilon _{AB} \; 
\bar{q}^b_{\mathrm{ch} \, A} (x) \; \gamma _5 \; q^c_B (x) \, , 
\label{eq:diq}
\en 
where $q^b_{\mathrm{ch} \, A} (x)$ are the charge conjugated quark 
fields, $A,B = 1 \ldots 2$ are flavor- and $a,b,c = 1 \ldots 3$ are 
color indices, respectively. The Effective Action, which should describe 
physics at the transition scale, is a SU(3) pure gauge theory supplemented 
with a scalar (Higgs, diquark) field which belongs to the 
the fundamental representation of the gauge group. 
The Lattice Effective Theory is modeled by the action 
\bea 
S &=& - \frac{\beta }{3} \sum _{\mu <\nu, x} \tr \; U_\mu (x) U_\nu (x+\mu) 
U^\dagger _\mu (x+\nu ) U^\dagger _\nu (x )  
\label{eq:action} \\ 
&+& \sum _x \phi ^a(x) \phi ^a(x) \; - \; \kappa \sum _{\mu, x} 
\phi (x) U_\mu (x) \phi ^\dagger (x+\mu ) \; - \; \mathrm{h.c.} 
\; + \; \sum _x \lambda [\phi ^2(x) ]^2 \; .
\nonumber 
\ena 
Thereby, the gluon degrees of freedom are encoded by the link 
fields $U_\mu (x)$, 
$\beta $ is the usual prefactor of the Wilson action, which largely describes 
the gluon-dynamics, and $\kappa $ is the Higgs hopping parameter which 
is related to the (bare) Higgs mass $m$ by 
\be
m^2 a^2\; = \; 1 \; - \; 8 \, \kappa \; , 
\label{eq:bec}
\en
where $a$ is the lattice spacing. Turning off the gluon interaction 
$(U_\mu =1)$, a Bose-Einstein condensate (BEC) is formed for $\kappa > 1/8$. 
Thereby, the BEC regime is stabilized by the Higgs quartic term 
$(\lambda >0)$.

\subsection{ The $SU(3)$ Higgs mechanism and residual confinement } 

Confinement effects in the gauged SU(2) Higgs model were extensively 
studied in~\cite{Langfeld:2001he,Langfeld:2002ic,Bertle:2003pj}. 
Here, the SU(3) Higgs system will be investigated for the first time. 

\vskip 0.3cm
For a sufficiently large Higgs hopping parameter $\kappa $, 
we expect that the system passes into the phase of condensed 
diquarks. In order to realize the formation of a scalar 
expectation value, we are forced to fix the gauge degree of freedom: 
Since a gauge transformation $\Omega (x) \in \,$SU(3) acts on the fields 
as 
$$
U^\Omega _\mu (x) \; = \; \Omega (x) \, U _\mu (x) \, \Omega ^\dagger 
(x+\mu ) \; , \hbo \phi ^\Omega (x) \; = \;  \Omega (x) \, \phi (x) \; , 
$$ 
any residual gauge degree of freedom would wipe out the expectation 
value $\langle \phi (x) \rangle $. Here we choose the Minimal Landau 
Gauge, i.e., 
\be 
\sum _{\mu ,x } \; \tr \; U^\Omega _\mu (x) \; \stackrel{\Omega }{ 
\longrightarrow } \mathrm{maximal} \; . 
\label{eq:gauge} 
\en 
Note that the gauge constraint (\ref{eq:gauge}) leaves a global gauge 
transformation $\Omega (x) = \Omega $ unfixed. The spontaneous breaking 
of this residual global gauge symmetry is signaled by 
a non-vanishing value $\langle \phi ^\Omega (x) \rangle $ and marks 
the occurrence of the Higgs phase~\cite{Langfeld:2002ic}. 

\vskip 0.3cm
Which fields gain a mass due to the formation of the BEC of diquarks? 
What is the fate of the Would-Be Goldstone bosons? What is different 
for SU(3) compared with the familiar SU(2) Higgs mechanism? 
In order to answer these questions, let us invoke a semi-classical 
approach for the moment. Thereby, the scalar field is decomposed into 
a classical part and fluctuations, $\phi ^\Omega (x) \; = \; 
\phi _c + \varphi (x) $. After (minimal) Landau gauge fixing, we may 
choose without a loss of generality
\be 
\phi _c \; = \; v \; (1,0,0)^T \; ; \hbo 
\mathrm{SU(3)} \longrightarrow \mathrm{SU(2)} \; . 
\label{eq:higgs}
\en 
This implies that only a part of the global SU(3) color group is 
broken by the condensate, and that a SU(2) color symmetry remains 
intact. 
From the Higgs kinetic term, we detect the masses of the gluon fields 
$A_\mu ^a$, $a=1\ldots 8 $, i.e., 
$$ 
\left[D_\mu \phi (x) \right]^\dagger D_\mu \phi (x) \; = \; 
\ldots \; + \; A_\mu ^a (x) M^{ab} A_\mu ^b(x) \; , \hbo 
M^{ab} = \frac{1}{2} \phi_c^\dagger \{t^a,t^b\} \phi _c \; , 
$$
where $D_\mu $ is the gauge covariant derivative, and $t^a$, $a=1 \ldots 8$ 
are the generators of the SU(3) algebra. Using (\ref{eq:higgs}), a 
direct calculation of the mass matrix $M^{ab}$ reveals that 
$$
A_\mu ^1 , \; A_\mu ^2 , \; A_\mu ^3 + \sqrt{3} \, A^8_\mu, \; 
A_\mu ^4 , \; A_\mu ^5 : \; \mathrm{massive} ; \; \; \; 
A_\mu ^6 , \; A_\mu ^7 , \; A_\mu ^3 - \sqrt{3} \, A^8_\mu : \; 
\mathrm{massless} . 
$$ 
Unless in the case of SU(2), there are not enough Higgs fields 
to give a mass to {\it all } gluons. As expected, the gluons corresponding 
to the unbroken global SU(2) color symmetry remain massless. 
The interesting question which solely arises in the context of SU(3) (and 
which will be partially answered below) is 
whether color charges which transform under the invariant SU(2) 
subgroup are still confined. Since the diquark field $\phi ^1$ 
is built up from quarks of color 2 and 3 (see \ref{eq:diq}), 
the integrity of  $\phi ^1$ as point-like particle would be preserved 
by confining effects throughout the BEC transition. 

\vskip 0.3cm
In order to explore the phase diagram of the Lattice Effective Theory 
of gluons and diquarks as function of $\beta $ and $\kappa $ 
($\lambda $ will be a given number in the studies below), 
we need an order parameter which detects the BEC phase. 
After installing the gauge condition (\ref{eq:gauge}), one might think 
to use $\langle \phi _c \rangle $. The point is that 
in an ergodic lattice simulation each lattice configurations would 
generate a different direction for $\phi _c$ implying that 
$\langle \phi _c \rangle =0$ by virtue of the residual {\it global } 
gauge degree of freedom. Let us define 
\be 
v^a \; = \;  \frac{1}{N_x} \sum _x \phi ^{\Omega \; a} (x)   \; , 
\label{eq:v} 
\en 
where $N_x$ is the number of space-time points. 
The crucial observation is that in the BEC phase the fields 
$\phi ^\Omega (x) $ are (almost) uniquely oriented throughout space-time,
i.e., $v^2 = {\cal O}(1)$, while in the color unbroken phase 
$v^2 \approx 0$. This suggests to use 
\be 
\Phi ^2 \; = \;  \left\langle 
\frac{1}{N^2_x} \sum _a \; \left[ \sum _x \phi ^{\Omega \; a} (x) \right] 
\;  \left[ \sum _y \phi ^{\Omega \; a} (y) \right] \right\rangle 
\label{eq:cond}
\en 
as the Litmus paper for the BEC transition. In the (global) color 
unbroken phase, the correlation length $\xi $ of the gauged scalar fields 
is defined from the (disconnected) Green function by 
\be
\langle  \phi ^{\Omega \; a} (x)  \phi ^{\Omega \; a} (y) 
\rangle \propto \exp \left\{ -  \vert x-y \vert / \xi 
\right\} \; , \hbo \xi \; \mathrm{finite} \; . 
\label{eq:corr}
\en
We therefore find that $\Phi ^2 $ vanishes in the infinite 
volume limit, i.e., 
$$
\lim _{N_x \to \infty } \Phi^2 \; \approx \; \lim _{N_x \to \infty } 
\xi/N_x \; = \; 0 \; , 
$$
while $ \Phi^2 = {\cal O} (1)$ in the BEC phase. 

\subsection{Numerical results } 

\begin{figure}
\caption{\label{fig:1} The diquark condensate defined in (\ref{eq:cond})
as function of the Higgs hopping parameter $\kappa $ for several 
values $\beta $ (for a definition of the parameters see (\ref{eq:action})). 
The colored symbols indicate the positions in parameter space where 
finite density simulations will be carried out. 
}
\includegraphics[height=.25\textheight]{conf6_phi}
\includegraphics[height=.25\textheight]{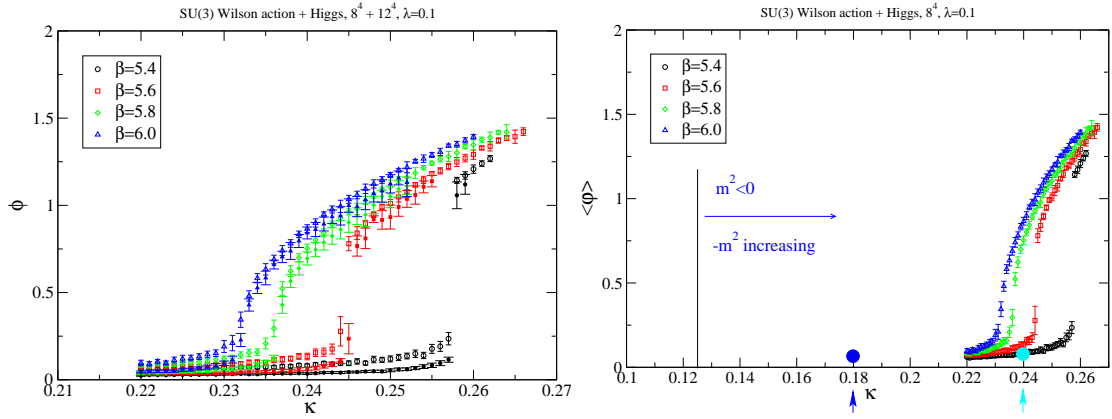}
\end{figure}

The Lattice Effective Theory, corresponding to the action 
(\ref{eq:action}), can be simulated on a computer using 
a generalized version of the Cabibbo Marinari algorithm. 
Micro-canonical reflections concerning both, the scalar fields and 
the link fields, are employed to reduce autocorrelations. 
In this first investigation, neither a scaling analysis nor 
a study of the line of constant physics is pursued. The aim of the 
simulations was to reveal the underlying physics at a qualitative 
level. The numerical results were obtained for $\lambda  = 0.1$ on 
$8^4$ and $12^4$ lattices. Landau gauge, see (\ref{eq:gauge}), is 
implemented using a standard iteration over-relaxation procedure. 

\vskip 0.3cm
The findings for the ``diquark condensate'' $\Phi $ (\ref{eq:cond}) 
are summarized in figure~\ref{fig:1}. The open symbols have been 
obtained on a $8^4$ lattice, while the full symbols correspond 
to a $12^4$ lattice. 

\vskip 0.3cm
We find that diquark 
condensation sets in if $\kappa $ exceeds a critical value. 
This finding is a highly non-trivial result for the following reason: 
Landau gauge fixing is performed at the level of the link fields 
and implies the maximization of the gauge functional (\ref{eq:gauge}). 
In particular the iterative procedure used here only locates a 
{\it local } maximum, and repeating the maximization on the 
same link configurations generically yields different gauge transformations 
$\Omega (x)$, each of which generates a different Gribov copy of the links 
and scalar fields, respectively.  This implies that an 
average over the Gribov copies within the first Gribov horizon is 
performed when the expectation value $\Phi $ (\ref{eq:cond}) is 
calculated. Since the scalar field transforms homogeneously, the 
first guess would be that the correlation of the gauged scalar fields 
is destroyed by the average over the first Gribov horizon. 
The non-trivial result is that this is not the case: 
The scalar correlation length $\xi $ 
(\ref{eq:corr}) is {\it insensitive} to the Gribov noise. 

\vskip 0.3cm
Having in mind that in the case without gluonic interactions 
Bose-Einstein condensation sets in for $\kappa > 0.125 $, see 
(\ref{eq:bec}), we here observe that the onset of the condensation 
is postponed to much larger values of $\kappa $ due to gluonic 
interactions. A coexistence of the hadronic phase and the 
diquark BEC phase is not observed. 

\vskip 0.3cm
For $\beta $ values as large as $5.8$ (and for a $8^4$ lattice), the system 
is in the deconfinement phase by virtue of temperature and volume 
effects. Figure \ref{fig:1} shows that the BEC transition changes 
form first order to second order (or higher) when $\beta $ is increased. 
A possible explanation for this observation is: 
If the gluons are deconfined due to volume effects, only five 
of them acquire a mass by virtue of the SU(3) Higgs 
mechanism. Three gluons remain massless and give rise 
to the critical phenomenon with infinite correlation length. 
At small $\beta $ values, the system is in the ``hadronic'' phase 
and the gluons possess a mass gap. If confinement persists 
for color states corresponding to the unbroken SU(2) subgroup, 
the three SU(2) gluons possess a mass gap due to confinement. 
The remaining five gluons are massive because of the Higgs 
mechanism. This would imply that there is no massless excitation 
which could give rise to a second order transition. 
This line of arguments favors the picture that color states 
of the residual SU(2) subgroup are still confined after the BEC 
transition.

\section{ Finite density deconfinement transition } 

\subsection{Lattice gauge theory  results  } 

The generic approach to Yang-Mills thermodynamics at finite baryon 
densities is based on the introduction of a non-zero chemical potential. 
In the case of a SU(2) gauge group, the fermion determinant is real and 
can be included in the probabilistic measure. Numerical simulations can 
be performed by using standard algorithms, although this numerical 
approach consumes a lot of computer time due to the non-local nature 
of the action~\cite{Hands:1999md}. In the case of a SU(3) gauge group, 
the fermion determinant acquires imaginary parts for a non-vanishing 
chemical potential and cannot be considered to be part of the probabilistic 
measure. The most prominent example to circumvent this conceptual 
difficulty considers the fermion determinant as part of the correlation 
function to be calculated. Thereby, the probabilistic measure of 
zero-density Yang-Mills theory is used to generate the gauge field 
configurations. However, it turns out that this approach suffers from 
the so-called ''overlap'' problem implying that for realistic lattice 
sizes an unrealistic number of Monte-Carlo steps is necessary to 
achieve reliable results~\cite{Barbour:1998jx}. 

\vskip0.3cm
At the present stage, the scope of lattice QCD simulations is 
limited to the regime of small baryon densities. Two approaches 
have been proven to be fruitful: (i) the approach based upon a 
Taylor expansion with respect to the chemical potential $\mu $ around 
$\mu =0$~\cite{Ejiri:2003dc}; (ii) the method employing 
simulations at imaginary chemical potential and finally seeking 
a continuation to real chemical potential~\cite{deForcrand:2002ci}. 
Finally, I would like to mention a recently proposed technique 
where multi-parameter re-weighting is used in order to reduce the severeness 
of the overlap problem~\cite{Fodor:2004nz}. 

\vskip0.3cm
It is fair to say that a direct lattice study of the QCD phase transition 
at intermediate baryon densities (and small temperature) is not feasible  at 
the present level of investigations. 

\subsection{ A change of view } 

The solution of the sign/overlap problem in finite density 
lattice QCD probably requires new type of algorithms such as cluster 
algorithms or D-theory~\cite{Alford:2001ug,Chandrasekharan:1996ih}.
Here, I would like to suggest to change the question of interest in 
a way which makes the problem solvable and which nevertheless 
sheds light onto the region of the QCD phase diagram where 
the finite density deconfinement transition takes place. 

\vskip 0.3cm 
In order to put my proposal into the proper context, let me 
briefly review the origin of the sign problem. The standard question 
which we used to ask is: what is the ground state energy of the system 
if we only consider field configurations which are eigen states 
of the particle number operator $\hat{N}$: 
\be 
E(B ) \; = \; \langle \phi \vert \hat{H} \vert \phi \rangle \; , \hbo 
\hat{N} \; \vert \phi \rangle \; = \; B \; \vert \phi \rangle \; ,  
\label{eq:en_st} 
\en 
where $ \hat{H} $ is the Hamilton operator. A conversion of the latter 
formulation to a functional integral setup usually involves the 
introduction of a chemical potential and generically leads to the 
sign problem. 

\vskip 0.3cm 
Here, I propose to consider, instead of (\ref{eq:en_st}), the ground state 
energy where the fields {\it on average } possess a given 
particle number, i.e., 
\be 
E(B) \; = \; \langle \phi \vert \hat{H} \vert \phi \rangle \; , \hbo 
\langle \phi \; \vert \hat{N} \; \vert \phi \rangle \; = \; B \; . 
\label{eq:en_cl} 
\en 
This approach certainly disregards certain features of the multi-fermion 
system; the hope is, however, that the approach sketches the deconfinement 
transition at high densities at least qualitatively. 
It is well known how to formulate the approach (\ref{eq:en_cl}) in 
the functional integral language~\cite{coleman}. 
The quantity of interest is the 
effective action $\Gamma $, which originates from the partition 
function by means of a Legendre transformation: 
\bea 
Z[\mu ] &=& \int {\cal D}\phi \; \exp \left\{ -S \; + \; 
\int d^4x \;  \mu (x) \; \rho (x) \right\}  \; , 
\label{eq:e_act_a} \\ 
\Gamma [\rho _c] &=& - \ln  Z[\mu ] \; + \; \int d^4x \; \mu (x) \; 
\rho _c (x) \; ,  \hbo 
\rho_c (x) \; = \; \frac{ \delta \ln  Z[\mu ] }{ \delta \mu (x) } \; ,  
\label{eq:e_act} 
\ena 
where $\rho (x) $ is the zeroth component of U(1) Noether current 
corresponding to conserved baryon charge. Using a constant 
external source, i.e., $\mu (x) = \mu $, the effective action is 
turned into the effective potential for a constant baryon density $\rho _c$. 
The technical advantage of the approach to systems of finite 
(classical) density is that the additional factor $ exp 
\int d^4x \;  \mu (x) \; \rho (x) $ in (\ref{eq:e_act_a}) 
can be included to the action during the Monte-Carlo Updates leading to 
significant overlap with the finite density configurations.

\subsection{ LET study of the finite density transition } 

Let me stress that the classical density approach, outlined in the 
previous subsection, is applicable to lattice QCD. 
Nevertheless, such simulations involve the inclusion of dynamical quarks 
and are very time consuming. For this reason, we will here explore 
the method resorting to the Lattice Effective Theory (LET) discussed in the 
previous section. 

\vskip 0.3cm 
In the present case, the baryon density is solely supported by the 
diquarks. The corresponding Noether density $\rho (x) $ is given by the 
zeroth component of the current 
$$ 
j_\mu (x) \; = \; \frac{1}{2i} \left[ \phi ^\dagger (x) 
\, U_\mu (x) \, \phi (x+\mu ) \; - \; \phi ^\dagger (x+\mu ) \, 
 U^\dagger _\mu (x) \, \phi (x) \; \right] \; . 
$$
The action of the LET is then given by 
\be 
S_\mathrm{den} \; = \; S \; - \; \mu \sum _x \rho (x) \; , 
\en
where the zero density part $S$ is defined in (\ref{eq:action}). 
It is still feasible to modify the Cabibbo Marinari algorithm to include 
the finite $\mu $ term making the lattice simulation straightforward. 

\begin{figure}
\caption{\label{fig:2} Left Panel: The expectation value of the baryonic 
Noether current as function of the external source $\mu $. The blue 
crosses indicate the parameters with which the probability distribution 
of $\rho $ will be studied below. Right panel: 
Illustration of the baryonic matter contributing to $\rho _c$. 
}
\includegraphics[height=.25\textheight]{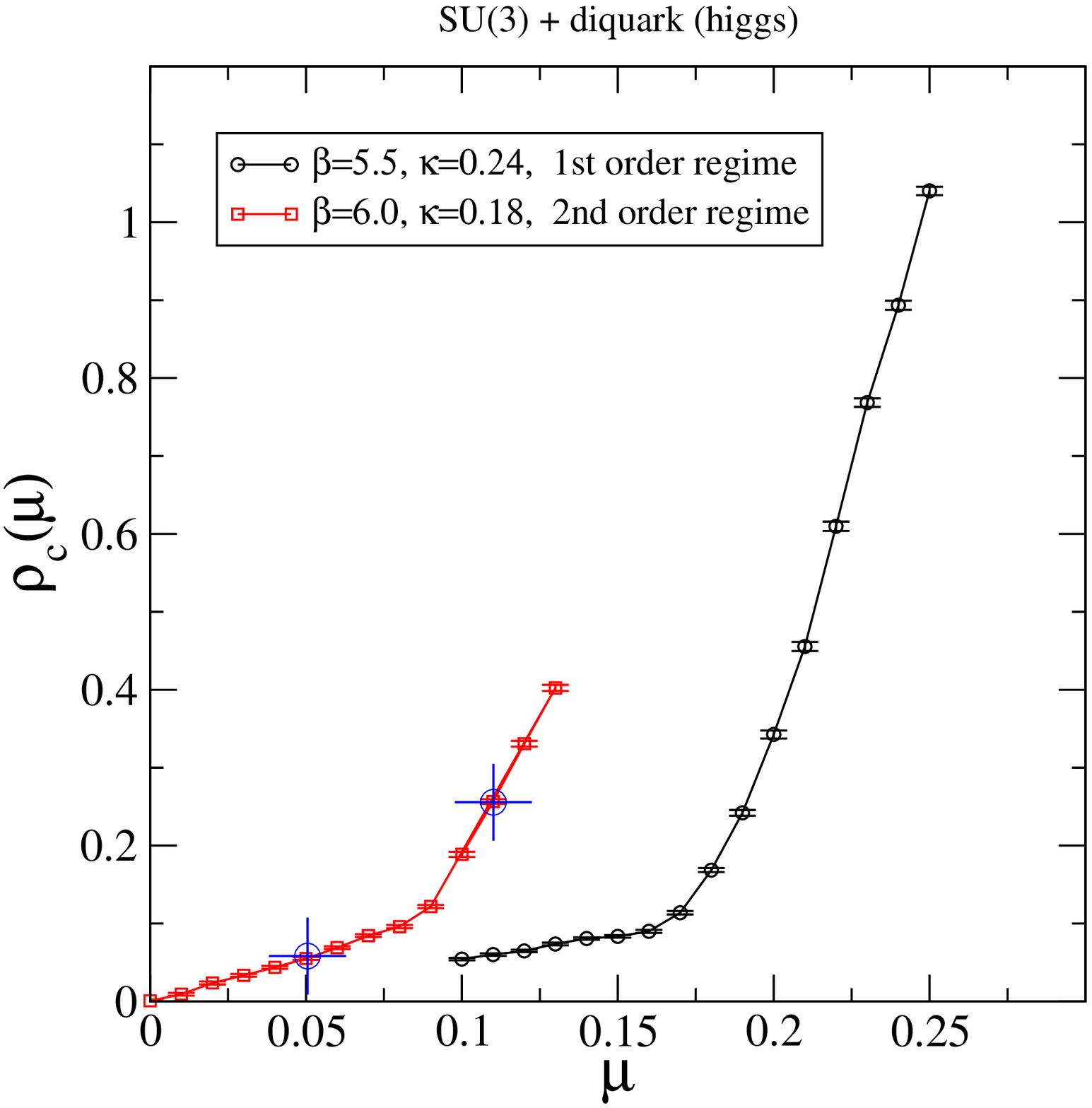}
\includegraphics[height=.25\textheight]{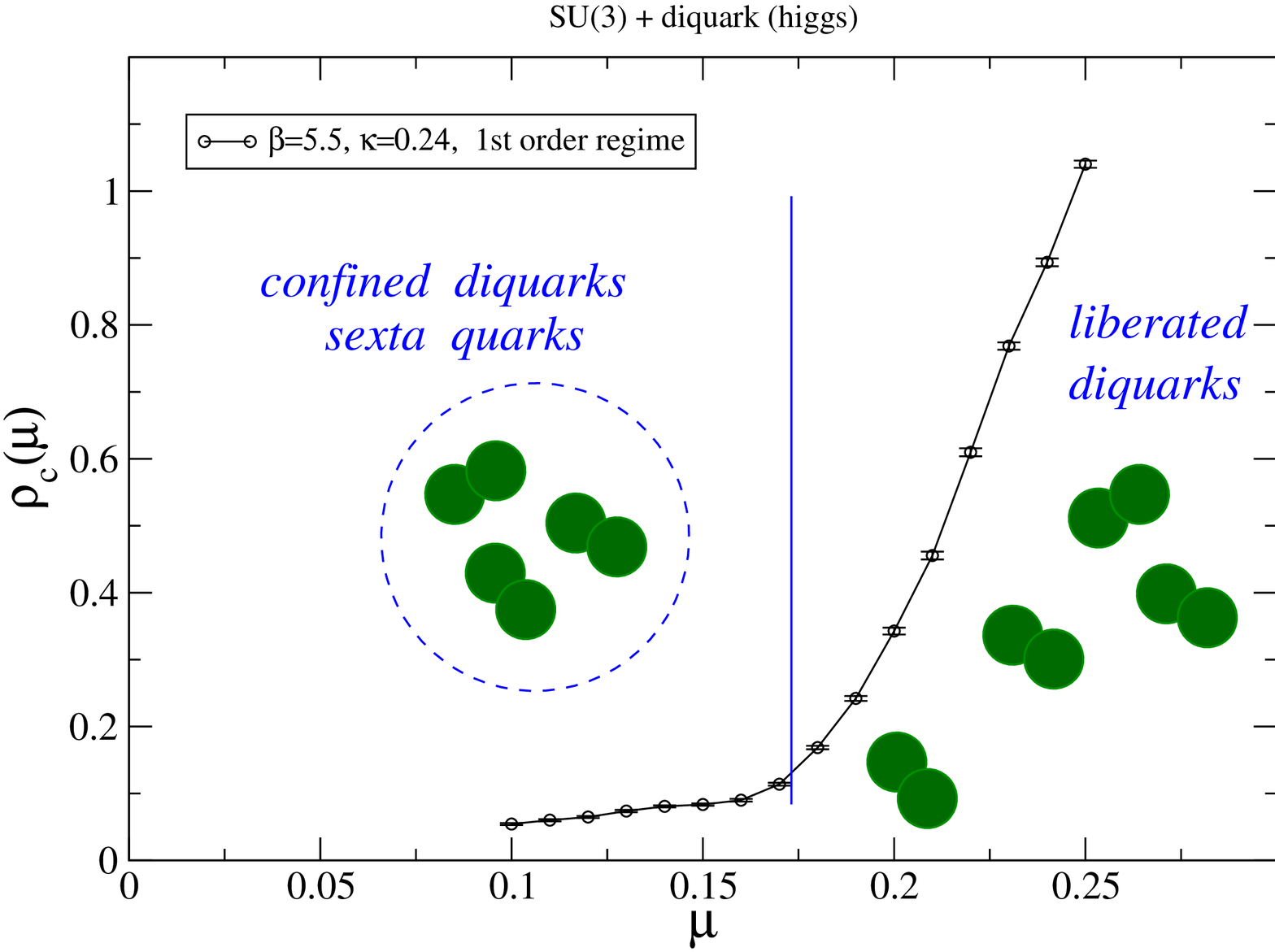}
\end{figure}

\vskip 0.3cm
For the actual simulation, two regimes are of particular interest: (i) 
The large volume regime, where the gluons which belong to the residual 
unbroken SU(2) subgroup are confined. (ii) The small volume 
regime, where the Higgs transition (as function of the Higgs hopping 
parameter $\kappa $) is 2nd order. 
Here, I used the parameters (see figure \ref{fig:1}, right panel 
for an illustration) 
\bea 
\beta \; = \; 5.5 \; ,  \;  \;  \; \kappa &=&  0.24 \hbo 
\hbox{(large volume regime)} 
\nonumber \\ 
\beta \; = \; 6.0 \; ,  \;  \;  \; \kappa &=&  0.18 \hbo 
\hbox{(small volume regime)} \; . 
\nonumber 
\ena 
The results are presented in figure \ref{fig:2}. We observe that 
the behavior of the expectation value of the baryonic Noether current, 
$\rho_c=\langle \rho \rangle $, as function of the overlap enhancing factor 
$\mu $ is qualitatively the same for both scenarios: 
$\langle \rho \rangle $ increases linearly with $\mu $ 
until a critical value $\mu _c$ is reached. For $\mu > \mu_c$, 
the linear dependence continues with a bigger slope. There is 
an intuitive understanding for this behavior (for an illustration 
see figure \ref{fig:2}, right panel): At small $\mu $, we are in 
a ``confined phase'' where small color electric flux tubes 
connect static color sources until string breaking occurs at 
large distances. In this phase, the diquarks are bound to color 
singlet states which are in the present theory 3-diquark states 
or, equivalently, 
sexta-quark states. For $\mu > \mu_c $, the theory looses its confining 
capabilities; diquarks are no longer bound to color singlet sexta-quarks. 
The baryon number susceptibility has increased due to 
deconfinement. It is interesting to note that even the Higgs transition 
is of first order when the Higgs hopping term is varied, the 
finite density transition is more like a crossover which describes 
sexta-quarks dissolving into liberated diquarks. 

\begin{figure}
\caption{\label{fig:3} Probability distribution of the baryonic 
Noether density. 
}
\includegraphics[height=.4\textheight]{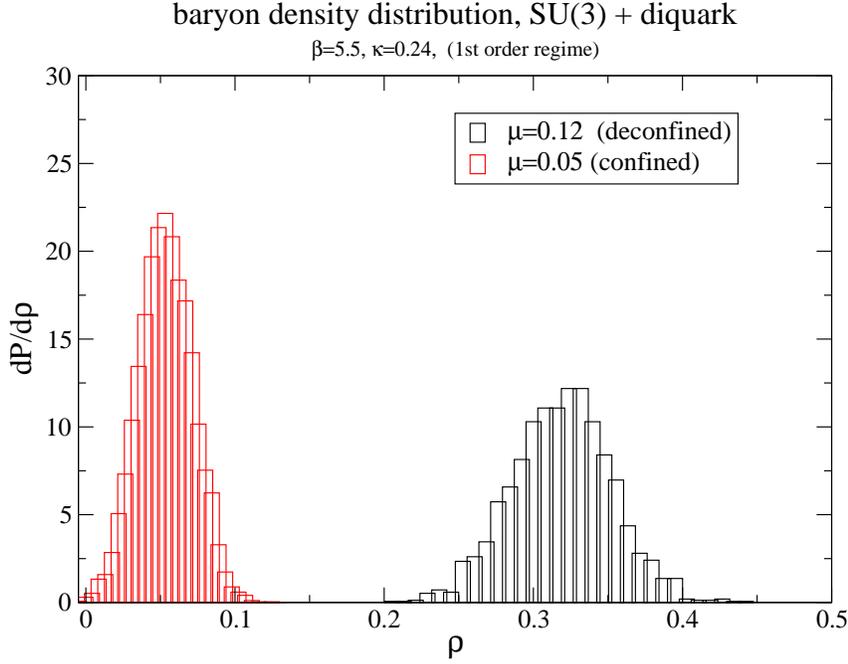}
\end{figure}

\vskip 0.3cm 
Let us discuss how a particular expectation value 
$\rho _c = \langle \rho \rangle $ is realized in a lattice simulation 
for a given external parameter $\mu $. For this purpose, we ask: How 
big is the probability, $P(\rho ) \; d\rho $, of finding $\rho $ in the 
interval $[\rho, \rho + d\rho [$ for an actual lattice configuration? 
The normalized distribution $dP/d\rho $ is shown in figure \ref{fig:3} 
for $\mu = 0.05 $ (confined phase) and for $\mu = 0.12 $ 
(deconfined phase). It is intuitive that the distribution is broader 
in the deconfined regime. Finally, I point out that simulations 
for a given value of $\rho _c$ can be done without to much of a 
loss of statics: In this case, we would confine us to configurations 
with  $\rho $ belonging to a bin where $dP/d\rho $ peaks. 

\section{Conclusions} 

In this paper, the proposal~\cite{Langfeld:2002ic} to describe 
the deconfinement phase transition at high baryon density 
by an effective theory of diquarks and gluons is thoroughly 
investigated by means of lattice simulations. The dynamics of the 
Lattice Effective Theory (LET) is dictated by the action of a gauged SU(3) 
Higgs model where the scalar Higgs field simulates the point-like diquark. 

\vskip 0.3cm 
In a first step the regime of vanishing baryon density is explored. 
For this purpose, a detailed lattice simulation of the SU(3) Yang-Mills 
theory with a scalar field in fundamental representation was performed for the 
first time. In the large volume limit a first order deconfinement transition 
occurs, if the Higgs hopping parameter exceeds a critical strength. 
The diquarks undergo Bose-Einstein condensation. It turns out that 
only five of the eight gluons acquire a mass by the SU(3) Higgs 
mechanism, and preliminary evidence was found that confinement 
with respect to the unbroken SU(2) subgroup is intact. 

\vskip 0.3cm 
In order to gain first insights into the regime of intermediate 
baryon densities, I here propose to perform lattice simulations 
at a fixed expectation value  $\rho _c$ of the baryonic Noether current. 
This procedure is outline for the LET designed above. I stress, however, 
that this approach is applicable to full QCD simulations. 
The practical benefit of this change of view onto the density 
regime is that severe overlap and sign problems are avoided. 
Overlap is ensured by an external parameter $\mu $ which couples to the 
Noether density. 

\vskip 0.3cm 
At small $\mu $, the linear rise of $\rho _c$ with increasing $\mu $ 
is due to the population of the vacuum with color singlet bound states 
consisting of three diquarks (sexta-quarks). Above a critical value, 
deconfinement occurs, and the vacuum is populated by liberated diquarks. 
The transition appears to be a crossover. A significant signal of 
deconfinement is only visible in the baryon  number susceptibility which 
rapidly increases at the transition. 

\vskip 0.3cm 
More realistic LETs also incorporate the ``valence quark'' and 
give rise to baryons as bound states of diquarks and valence quarks. 
The investigation of these LETs are left to future work.

\begin{theacknowledgments}
  This work is supported in parts by the Ministry of Science, Research 
  and the Arts of  Baden W\"urtemberg under Az: 24-7532.23-19-18/1 
  and by the {\it Virtual Institute } of the Helmholtz Association 
  no.~VH-VI-041. 
\end{theacknowledgments}


\bibliography{sample}

\end{document}